\documentclass[useAMS,usenatbib]{mn2e}

\usepackage[dvips]{graphicx}   
\usepackage{mathrsfs}
\usepackage{color}

\usepackage{natbib}
\def\reff@jnl#1{{\rm#1\/}}
\def\aj{\reff@jnl{AJ}}                  
\def\araa{\reff@jnl{ARA\&A}}            
\def\apj{\reff@jnl{ApJ}}                
\def\apjl{\reff@jnl{ApJ}}               
\def\apjs{\reff@jnl{ApJS}}              
\def\apss{\reff@jnl{Ap\&SS}}            
\def\aap{\reff@jnl{A\&A}}               
\def\aapr{\reff@jnl{A\&A~Rev.}}         
\def\aaps{\reff@jnl{A\&AS}}             
\def\mnras{\reff@jnl{MNRAS}}            
\def\prd{\reff@jnl{Phys.Rev.D}}         
\def\prl{\reff@jnl{Phys.Rev.Lett}}      
\def\pasp{\reff@jnl{PASP}}              
\def\pasj{\reff@jnl{PASJ}}              
\def\nat{\reff@jnl{Nature}}             

\usepackage{amssymb}           

\newcommand{\bd}{\begin{displaymath}}
\newcommand{\ed}{\end{displaymath}}
\newcommand{\be}{\begin{equation}}
\newcommand{\ee}{\end{equation}}
\newcommand{\beaa}{\begin{eqnarray*}}
\newcommand{\eeaa}{\end{eqnarray*}}
\newcommand{\bea}{\begin{eqnarray}}
\newcommand{\eea}{\end{eqnarray}}

\newcommand{\boldsymbol}[1]{\mbox{\boldmath{${#1}$}}}

\def\tdist{D_{\Delta t}}

\def\imVec{\boldsymbol{\theta}}
\def\im{\theta}
\def\imA{\theta_{\rm A}}
\def\imB{\theta_{\rm B}}
\def\imN{\theta_0}
\def\imAi{\theta_{\rm A1}}
\def\imBi{\theta_{\rm B1}}
\def\imAii{\theta_{\rm A2}}
\def\imBii{\theta_{\rm B2}}

\def\imMod{\vartheta}
\def\imModA{\vartheta_{\rm A}}
\def\imModB{\vartheta_{\rm B}}
\def\imModAi{\vartheta_{\rm A1}}
\def\imModBi{\vartheta_{\rm B1}}
\def\imModAii{\vartheta_{\rm A2}}
\def\imModBii{\vartheta_{\rm B2}}
\def\srVec{\boldsymbol{\beta}}
\def\sr{\beta}
\def\srS{\beta_{\rm s}}
\def\srSi{\beta_{\rm s1}}
\def\srSii{\beta_{\rm s2}}

\def\slope{\gamma'}
\def\thEin{\theta_{\rm E}}
\def\kext{\kappa_{\rm ext}}

\def\dtBA{\Delta t_{\rm BA}}
\def\dt{\Delta t}

\def\helens{{\rm HE}1104$-$1805}
\def\HST{\textit{HST}}

\def\dataVec{\boldsymbol{d}}

\def\esrVec{\boldsymbol{s}}

\def\parsVec{\boldsymbol{\eta}}

\def\agnVec{\boldsymbol{\nu}}

\def\eagnVec{\boldsymbol{a}}

\def\Bmat{\mathbfss{B}}
\def\Lmat{\mathbfss{L}}

\title[Cosmography from two-image lens systems]{Cosmography from
  two-image lens systems: overcoming the lens profile slope
  degeneracy}
\author[S. H. Suyu]
{S. H. Suyu$^{1,2}$\thanks{E-mail:suyu@physics.ucsb.edu} \\
$^{1}$Department of Physics, University of California,
  Santa Barbara, CA 93106-9530, USA \\
$^{2}$Kavli Institute for Particle Astrophysics and Cosmology, Stanford University, 452 Lomita Mall, Stanford, CA 94035-4085, USA}

\begin{document}

\date{Accepted ---; received ---; in original form \today}

\pagerange{\pageref{firstpage}--\pageref{lastpage}} \pubyear{2012}

\maketitle

\label{firstpage}


\begin{abstract}

  The time delays between the multiple images of a strong lens system, 
  together with a model of the lens mass distribution, allow a one-step
  measurement of a cosmological distance, namely, the ``time-delay
  distance'' of the lens ($\tdist$) that encodes cosmological
  information.  The time-delay distance depends
  sensitively on the radial profile slope of the lens mass distribution;
  consequently, the lens slope must be accurately constrained for
  cosmological studies.  We show that the slope cannot be constrained in
  two-image systems with single-component compact sources, whereas it
  can be constrained in systems with two-component sources provided the
  separation between the image components can be measured with milliarcsecond
  precisions, which is not feasible in most systems.  In contrast, we
  demonstrate that spatially extended images of the source galaxy in
  two-image systems break the radial slope degeneracy and allow 
  $\tdist$ to be measured with uncertainties of a few percent.  Deep
  and high-resolution imaging of the lens systems are needed to reveal
  the extended arcs, and stable point spread functions are required
  for our lens 
  modelling technique.  Two-image systems, no longer plagued by the radial
  profile slope degeneracy, would augment the sample of useful time-delay
  lenses by a factor of $\sim$$6$, providing substantial advances
  for cosmological studies.

\end{abstract}



\section{Introduction} 
\label{sec:intro}

Since the discovery of the accelerated expansion of the Universe
\citep{PerlmutterEtal99, RiessEtal98}, one of the key puzzles
in cosmology has been the nature of dark energy which was proposed to
explain the accelerated expansion.  Recent studies based on various
cosmological probes including the cosmic microwave background (CMB), 
supernovae, baryon acoustic
oscillations, galaxy clusters, weak lensing, and gravitational lens
time delays have shown that the Universe is consistent with dark
energy being described by a cosmological constant $\Lambda$
\citep[e.g.,][]{KomatsuEtal11, ConleyEtal11, SuzukiEtal11, ReidEtal10,
  BlakeEtal11, MantzEtal10, SehgalEtal11, SchrabbackEtal10, SuyuEtal10}.
Nonetheless, \citet{Linder10a} showed that current data provide little
constraints on the properties and time evolution of dark energy when
one relaxes the assumption that the dark energy equation of state,
$w$, is constant (where $w=-1$ corresponds to the cosmological
constant).  To understand the nature of dark energy, a synergy of
future observations of independent cosmological probes (to overcome
the systematic effects in each approach), coupled with theoretical
investigations of dark energy models, is needed.

In this paper, we focus on a particular cosmological probe:
gravitational time delays in strong lens systems.  By measuring the
time delay(s) between the multiple images and modelling the mass
distribution of the lens galaxy, one can infer the ``time-delay
distance'', $\tdist$, to the lens system.  This distance, which is a
combination of angular diameter distances, is primarily sensitive to
the Hubble constant ($\tdist \propto H_0^{-1}$) but also depends on
other cosmological parameters such as $w$.  An accurate measurement of
the Hubble constant with uncertainties better than a few percent
provides the single most useful complement to results of the CMB for
dark energy studies \citep[e.g.,][]{Hu05, RiessEtal09, RiessEtal11}.
Furthermore, time-delay lenses are also highly complementary to
supernovae for determining the dark energy equation of state
\citep{Linder11}.

 \citet{SuyuEtal10} showed that
high-quality data of a four-image system allowed accurate lens mass
modelling which yielded competitive cosmological constraints.  On the
other hand, analyses of two-image systems, which have significantly
fewer time-delay and positional constraints on the mass model than
four-image systems, are often plagued by model degeneracies and thus
require model assumptions that may not be fully justified
\citep[e.g.,][]{BurudEtal02b, JakobssonEtal05, ParaficzEtal09}.
Ameliorating the shortcomings of two-image systems would provide
significant advances to time-delay cosmography since there are
currently more two-image systems than four-image systems
\citep[e.g.,][]{Oguri07} and future large-scale surveys expect to
discover about 6 times more two-image systems than four-image systems
\citep[e.g.,][]{OguriMarshall10}.

One of the main lens model limitations is due to the lens radial
profile degeneracy: for a power-law mass distribution with
three-dimensional density $\rho\propto r^{-\slope}$, there is a strong
degeneracy between the radial slope $\slope$ and $\tdist$.  While
studies of large strong lens samples from the Sloan Lens ACS
Survey (SLACS) indicate that lenses are well described by a
power law with $\slope\sim2.1$ \citep[e.g.,][]{KoopmansEtal09,
  AugerEtal10, BarnabeEtal11}, there is an intrinsic scatter in the
slope of $\sim0.15$.  Furthermore, studies of higher redshift lens
galaxies in the Strong Lensing Legacy Survey (SL2S) and the BOSS
Emission-Line Lens Survey (BELLS) find an evolution in the lens
profile slope where galaxies at higher redshift have shallower slopes
\citep{RuffEtal11, BoltonEtal12}.  Both the intrinsic scatter and the
evolution in the slope impact the $\tdist$ measurement.

\citet{WittEtal00} considered the time delays of power-law lens models
with arbitrary angular structure, and showed that for non-isothermal
mass distributions ($\slope\neq2$), the time delay depends in general
on $\tdist$, the image positions, $\slope$ and the source position (or
the lens potential).  \citet{Wucknitz02} investigated further the
power-law lens potentials with an additional external shear, and
derived $\tdist \propto \frac{3-\slope}{\slope-1}$ for fixed external
shear (where we have converted the notation from $\beta\equiv3-\slope$
and $H_0$).  This scaling is exact for power-law models with
  flexible angular structures that can fit perfectly to the
  observables, and is only approximate for elliptical models due to
  indirect dependencies of $\tdist$ on $\slope$ through, for example,
  the modelled source position.   Instead of parametrising in
terms of mainly the slope of the lens mass distribution,
\citet{Kochanek02} showed that the time delays primarily depend on the
average surface mass density in the annulus between the images
$\langle\kappa\rangle$, in addition to $\tdist$ and the image
positions.  When
expressed in terms of $\langle\kappa\rangle$, the correction to the
time delays from different values of $\slope$ is small.  The strong 
dependence of the time delays on the slope is incorporated indirectly
through $\langle\kappa\rangle$.  To highlight the full (both direct
and indirect) dependence of $\tdist$ on $\slope$ and its impact on
cosmography, we consider in the first part of the paper spherical
power-law models.  We also illustrate how one might constrain $\slope$
with more lensing data than just the image positions from a single
source, such as multiple compact source components or spatially
extended sources.

By using the extended images of the lensed source in optical
 or near-infrared (NIR) wavelengths to model both the 
lens mass distribution and the source surface brightness distribution,
studies have shown that the slope of the lens mass distribution can be
constrained with uncertainties of a few percent in the annulus covered
by the lensed images \citep[e.g.,][]{DyeWarren05, DyeEtal08,
  SuyuEtal10, VegettiEtal10}.  However, such studies focus mostly on
four-image systems, and the use of extended two-image systems for
cosmography has not been examined in detail.  Furthermore, the lens systems that
have been modelled so far using extended images in the
  optical/NIR wavelengths have relatively smooth 
variations in the image surface brightness.  In contrast, the lensed
sources in time-delay lenses typically have active galactic nuclei
(AGNs) that are much brighter than the AGN host galaxies and thus
require new modelling techniques to account for the large dynamical
range in surface brightness. 

The paper is organised as follows.  In Section \ref{sec:cosmo}, we
briefly review the method of gravitational lens time delays for
cosmography.  In Section \ref{sec:SPL}, we consider a spherical
power-law model to illustrate the degeneracy between $\tdist$ and
$\slope$ and how one would break the degeneracy in principle.  We
simulate observations of two-image lens systems with spatially
extended source galaxies in Section \ref{sec:sim}, and model these
systems to test the recovery of $\tdist$ in Section \ref{sec:Mod}.
Conclusions of our results are in Section \ref{sec:conclude}.


\section{Cosmography from gravitational lens time delays}
\label{sec:cosmo}

In this section, we give a brief overview of using strong lens systems
with measured time delays between the multiple images to study
cosmology.  More details on the subject can be found in, e.g.,
\citet{SchneiderEtal06} and \citet{Treu10}.  Readers familiar with
time-delay lenses may wish to proceed directly to Section
\ref{sec:SPL}.

According to Fermat's principle, the multiple images in a strong lens
system appear at locations where the travel times of the light paths
are extrema or saddles.  The excess time delay of an image at angular
position $\imVec=(\im_1,\im_2)$ with corresponding source position
$\srVec=(\sr_1,\sr_2)$ relative to the case of no lensing is 
\be
t(\imVec, \srVec) = \frac{\tdist}{c} \left[ \frac{(\imVec-\srVec)^2}{2}-\psi(\imVec) \right],
\ee
where $c$ is the speed of light, and $\tdist$ is the so-called
time-delay distance that is a  
combination of the angular diameter distance to the lens/deflector
($D_{\rm d}$) at redshift $z_{\rm d}$, to the source ($D_{\rm s}$), and
between the lens and the source ($D_{\rm ds}$):
\be
\tdist \equiv (1+z_{\rm d}) \frac{D_{\rm d} D_{\rm s}}{D_{\rm ds}}.
\ee
The lens potential $\psi(\imVec)$ is related to the
dimensionless surface mass density of the lens, $\kappa(\imVec)$, via 
\be
\label{eq:poisson}
\nabla^2 \psi(\imVec) = 2 \kappa(\imVec).
\ee

For systems which have sources with intensities that vary in time 
such as active 
galactic nuclei (AGNs), one can monitor the intensities of the lensed
images over time and measure the time delay, $\Delta t_{ij}$, between
the images at positions $\imVec_i$ and $\imVec_j$:
\bea
\nonumber \Delta t_{ij} \hspace{-0.2cm} &\equiv& \hspace{-0.2cm} t(\imVec_i,\srVec) -
t(\imVec_j,\srVec) \\
\label{eq:dt}
&=& \hspace{-0.2cm}
\frac{\tdist}{c} \left [  \frac{(\imVec_i-\srVec)^2}{2}-\psi(\imVec_i)
  - \frac{(\imVec_j-\srVec)^2}{2}+\psi(\imVec_j) \right ]. 
\eea

By using the image configuration and morphology, one can model the
mass distribution of the lens to determine the lens potential
$\psi(\imVec)$ and the unlensed source position $\srVec$.  Lens
systems with time delays can therefore be used to measure $\tdist$
via equation (\ref{eq:dt}) and constrain cosmological
models \citep[e.g.,][]{Refsdal64, Refsdal66, FadelyEtal10, SuyuEtal10}.  Since
lens and source redshifts typically span between $z_{\rm d}\sim 0.1-1$
and $z_{\rm s}\sim 1-3$, respectively, an advantage of using the
time-delay lenses for cosmography is that the method provides a
one-step physical measurement of a cosmological distance $\tdist$
independent of distance ladders.


\section{Spherical Power-Law Lens: Illustration of the
  radial profile slope degeneracy}
\label{sec:SPL}

Previous studies of gravitational lenses show that power-law mass
distributions provide adequate descriptions for lens galaxies 
\citep[e.g.,][]{KoopmansEtal09, SuyuEtal09, AugerEtal10, RuffEtal11,
  BarnabeEtal11}. In this section, we explore the properties of a
simple model: a spherical power-law mass distribution.  Despite its
simplicity, it clearly illustrates important parameter degeneracies,
particularly between the time-delay distance and the radial
slope.

\subsection{Surface mass density, lens potential and deflection angle}
\label{sec:SPL:lens}
A spherical power-law mass density distribution is of the form
\be
\rho(r) = \rho_0 r^{-\slope},
\ee
where $\rho_0$ is the normalisation, $\slope$ is the radial profile slope,
$r=\sqrt{x^2+y^2+z^2}=\sqrt{R^2+z^2}$ is the three-dimensional
radius, $R$ is the two-dimensional projected radius, and the $z$-axis
is chosen to be the line-of-sight direction.  An
isothermal mass distribution has $\slope=2$.  The
projected surface mass density along the line of sight is given by
\bea
\Sigma(R) &=& \int_{-\infty}^{+\infty} \rho(\sqrt{R^2+z^2})\,{\rm d}z \\
   &=& \frac{\rho_0 \sqrt{\pi}\,
     \Gamma(\frac{\slope-1}{2})}{\Gamma(\frac{\slope}{2})}
   \frac{1}{R^{\slope-1}}.
\eea

The dimensionless surface mass density (also known as the convergence)
for lensing studies is
\be
\kappa(\imMod) = \Sigma(\imMod D_{\rm d})/\Sigma_{\rm cr},
\ee
where $\imMod = R/D_{\rm d}$ and the critical surface mass density is
\be
\Sigma_{\rm cr}=\frac{c^2 D_{\rm s}}{4 \pi G D_{\rm d} D_{\rm ds}}.
\ee
For convenience, we rewrite the dimensionless surface mass density as
\be
\label{eq:kappa:thE}
\kappa(\imMod) = \frac{3-\slope}{2} \left[ \frac{\thEin}{\imMod} \right]^{\slope-1}
\ee
by subsuming the normalisation constants into $\thEin$, which is also
known as the ``Einstein radius''.  A point source that is located on
the optic axis ($z$-axis) extending from the observer through the centre of the lens
would be lensed into a ring with radius $\thEin$.  This circle 
marks the tangential critical curve of the lens system, and the mass
enclosed within the Einstein ring is
\bea
M_{\rm Ein} & = & \Sigma_{\rm cr} D_{\rm d}^2 \left ( 2\pi \int_{0}^{\thEin}
\kappa(\imMod) \imMod \rm{d}\imMod \right)\\
\label{eq:mEin}
 & = & \Sigma_{\rm cr} D_{\rm d}^2 \pi \thEin^2.
\eea
Note that the mass enclosed is directly dependent only on $\thEin$ and
not on $\slope$.  

The lens potential corresponding to the $\kappa(\imMod)$ in equation
(\ref{eq:kappa:thE}) can be obtained by solving equation 
(\ref{eq:poisson}) and is
\be
\psi(\imMod) = \frac{\thEin^2}{3-\slope} \left(\frac{\thEin}{\imMod}
\right)^{\slope-3}. 
\ee
The scaled deflection angle is
$\boldsymbol{\alpha}(\imVec)=\boldsymbol{\nabla}\psi(\imVec)$.  
For circularly symmetric surface mass densities
($\kappa(\imVec)=\kappa(\imMod)$ where
$\imVec=\imMod(\cos\,\varphi,\sin\,\varphi)$ in polar coordinates), the deflection
angle becomes
\be
\boldsymbol{\alpha}(\imVec)=\alpha(\imMod)\frac{\imVec}{\imMod} =
\alpha(\imMod) \hat{\imVec},
\ee
\citep[e.g.,][]{SchneiderEtal06}, and for the
$\kappa(\imMod)$ in equation (\ref{eq:kappa:thE}), we have
\be
\label{eq:alphaPL}
\alpha(\imMod) = \thEin \left( \frac{\thEin}{\imMod}
\right)^{\slope - 2}.
\ee

\subsection{Lens systems with single component sources}
\label{sec:SPL:1sr}

\begin{figure}
  \centering
   \includegraphics[width=0.35\textwidth, clip]{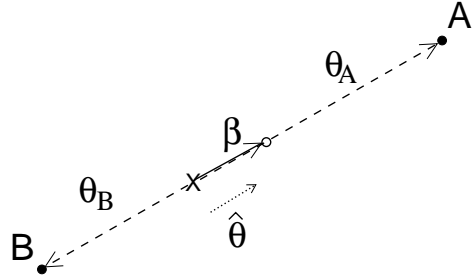}
  \caption{\label{fig:2ImSys} Two-image lens system from a
    circularly symmetric lens mass distribution.  The source (open
    circle) and images (filled circles) are collinear and are in the
    direction denoted by 
    $\hat{\imVec}$. The source position is at $\sr$ and the
    corresponding image positions are at $\imA$ and $\imB$. }
\end{figure}

Sources of time-delay lenses to date are AGNs with time-varying
intensities.  For practical purposes, the AGNs can be
treated as point sources.  The positions of the images of a source at
position $\srVec$ are obtained by solving the lens equation for
$\imVec$:
\be
\label{eq:lensEq}
\srVec = \imVec - \boldsymbol{\alpha}(\imVec).
\ee
In the case where the lens mass distribution is spherically symmetric,
the lens equation simplifies to 
\be
\sr = \im - \alpha(\imMod)\frac{\im}{\imMod},
\ee
where $\sr$ and $\im$ are measured from the lens centre and are
collinear. Note the distinction between $\im$ and $\imMod$, where
$\imMod = |\im|$.  For the deflection angle in equation 
(\ref{eq:alphaPL}), there are at most two non-central images of the
source that appear when $\sr \lesssim \thEin$.  Figure
\ref{fig:2ImSys} is an illustration of such generic two-image systems
from the spherical power-law model.  The cross marks the location of
the lens centre, and the two images are labelled by A and B.  The unit
vector $\hat{\imVec}$ indicate the direction of line where the source
(open circle)
and the corresponding images (filled circles) lie.  The source
position is at $\sr$, and 
the image positions are at $\imA$ and $\imB$, where the lens centre is
chosen to be the origin of the coordinates.

Given a two-image lens system as shown in Figure \ref{fig:2ImSys}, we
can use the image positions $\imA$ and $\imB$ to constrain the mass
distribution of the lens system.  In principle, the flux ratios of the
images can also be used, but in practise, the time variability and
delay between the images, substructure, microlensing and dust
extinction affect the flux ratio significantly, leading to large
uncertainties in the flux ratios\footnote{This is especially
    true in optical wavelengths, whereas in radio wavelengths,
    flux ratios are generally not affected by microlensing or
    extinction.}.  For simplicity, we consider only 
the image positions to probe the overall smooth component of lens mass
distribution.

Assuming that the lens mass centre can be determined based on its
light distribution, then the two image positions lead to the
following set of constraint equations:
\be
\label{eq:sys1sr}
\left\{ \begin{array}{l}
\srS = \imA - \alpha(\imModA)\frac{\imA}{\imModA} = \imA - \thEin \frac{\imA}{\imModA} \left( \frac{\thEin}{\imModA}\right)^{\slope-2} \\
\srS = \imB - \alpha(\imModB)\frac{\imB}{\imModB} = \imB - \thEin \frac{\imB}{\imModB} \left( \frac{\thEin}{\imModB}\right)^{\slope-2} \\
\end{array} \right.
\ee
where we have substituted in equation (\ref{eq:alphaPL}).  With three
model parameters ($\srS$, $\thEin$ and $\slope$) and two constraints,
the system of equations is underdetermined.  The paucity of
constraints from two-image systems explains why assumptions in the
lens mass distributions such as mass following light, spherical
symmetry and isothermality ($\slope=2$) were often necessary in
modelling two-image systems in the past \citep[e.g.,][]{VuissozEtal07,
  ParaficzEtal09}.  Here, we explore the dependence of
the model parameters and cosmological inferences on the slope $\slope$
in the range of 1.5 to 2.5, 
which reflect the spread in the lens slopes from the SLACS, SL2S
and BELLS samples \citep[e.g.,][]{KoopmansEtal09, AugerEtal10,
  BarnabeEtal11, RuffEtal11, BoltonEtal12}.  Since equations (\ref{eq:sys1sr})
cannot be solved analytically for generic values of $\slope$, we
consider six toy lens systems 
(I--VI) listed in Table \ref{tab:toySys} with different image
configurations. The image positions are expressed in units of $\imN$
that sets the overall size of the lens system.  We assume that
$\imModA\geq\imModB$ and 
$\imA>0$ (i.e., $\imB<0$) without loss of generality.

\begin{table}
\caption{Two-image toy systems}
\label{tab:toySys}
\begin{center}
\begin{tabular}[b]{ccc}
\hline
System & $\imA$ ($\imN$) & $\imB$ ($\imN$) \\
\hline
I & 1.0  & $-1.0$ \\
II & 1.2  & $-0.8$ \\
III & 1.2  & $-0.7$ \\
IV & 1.4  & $-0.7$ \\
V & 1.5  & $-0.3$ \\
VI & 1.8  & $-0.2$ \\
\hline
\end{tabular}
\end{center}
Notes.  Configuration of the six toy two-image lens systems.  Columns 2
and 3 are the positions of images A and B in units of $\imN$.
\end{table}

\begin{figure}
\centering
\includegraphics[width=0.47\textwidth, clip]{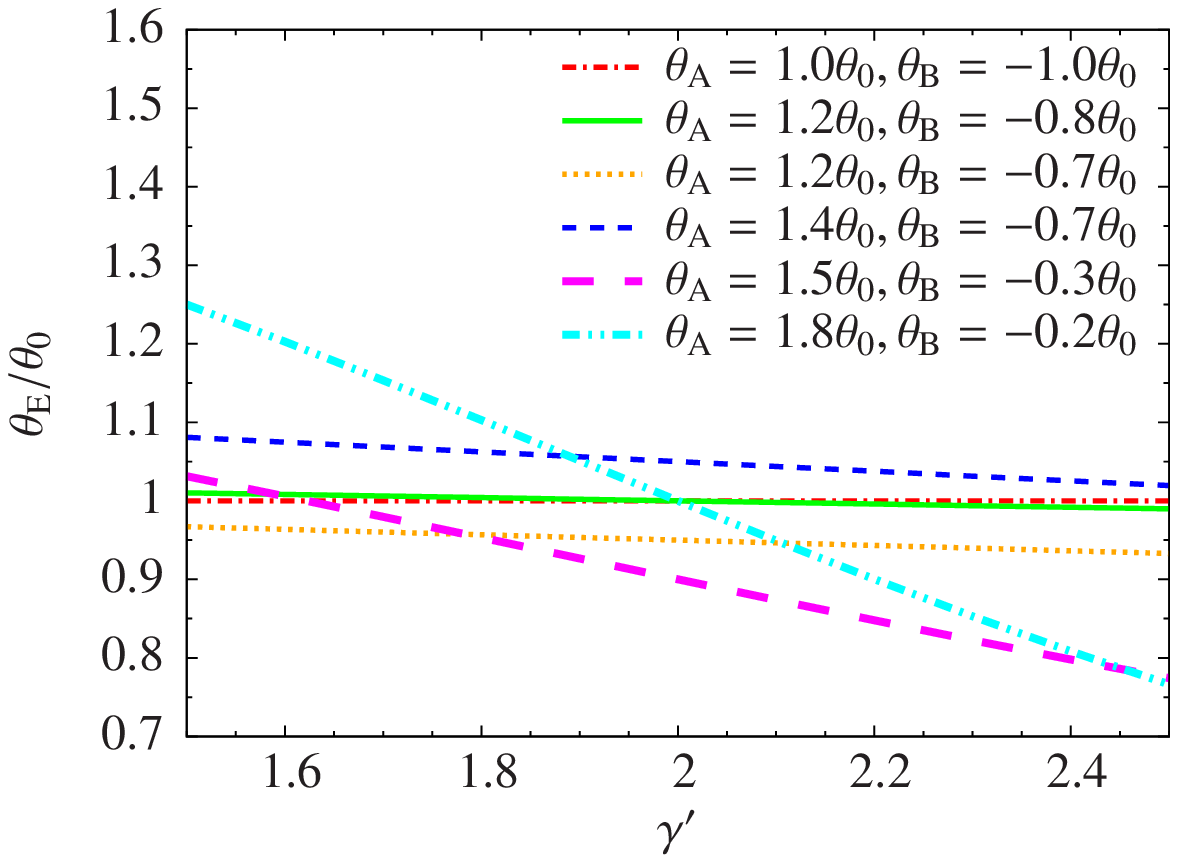}
\includegraphics[width=0.47\textwidth, clip]{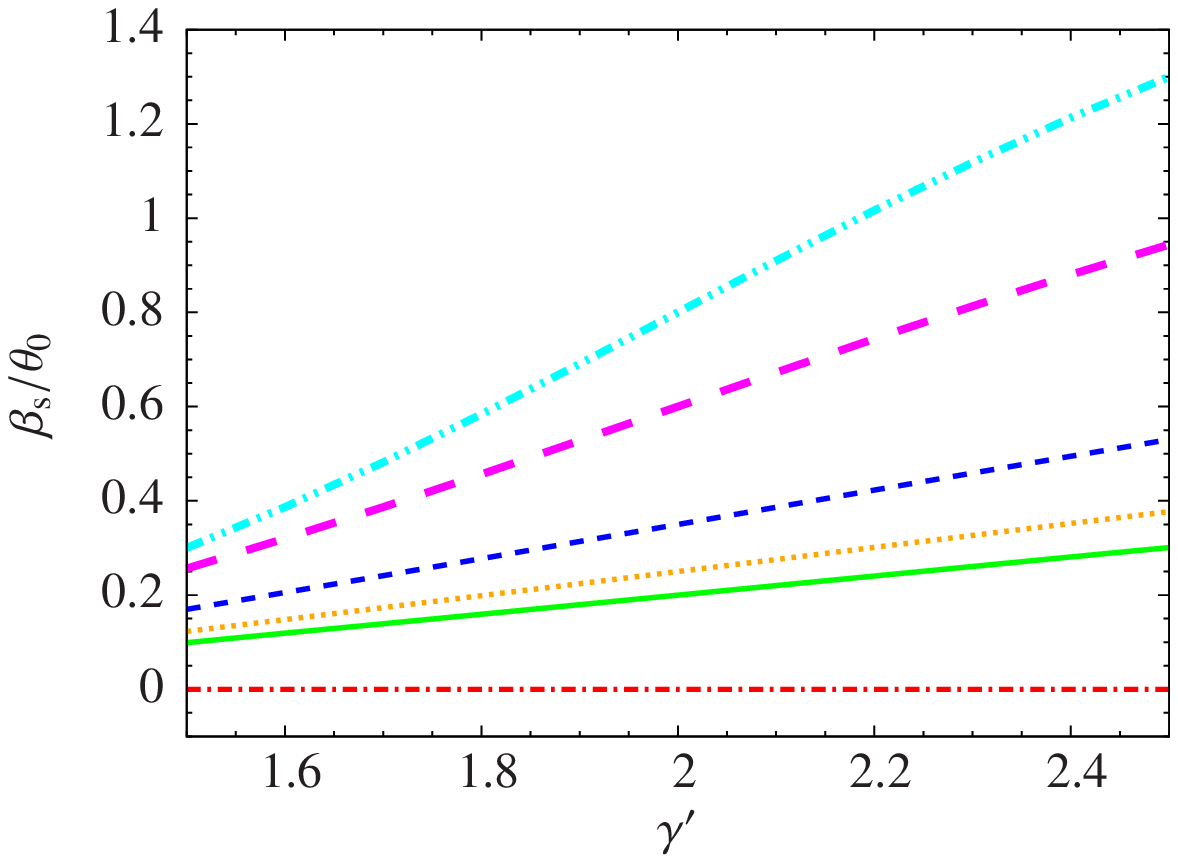}
\caption{\label{fig:thEandSr} The Einstein radius ($\thEin$) and source
  position ($\srS$) parameters for each of the six toy lens systems.
  For systems with images that are nearly symmetric with respect to
  the lens centre (i.e., $\imModA / \imModB \lesssim 2$), $\thEin$ is
  quite insensitive to lens profile slope $\slope$.}
\end{figure}

We show in Figure \ref{fig:thEandSr} values of $\thEin$ and $\srS$
that solve equations (\ref{eq:sys1sr}) for the range of slope
values.  For systems I to IV with nearly symmetric image configuration
($\imModA / \imModB \lesssim 2$), the value of $\thEin$ is quite
insensitive to $\slope$.  This implies that the mass enclosed within
$\thEin$ (which only has direct dependence on $\thEin$, as indicated
in equation (\ref{eq:mEin})) from strong lensing is accurate to within
$\sim$$5\%$ for these systems.  For the case with perfect symmetry
($\imModA = \imModB$) that has the source lensed into a ring, $\thEin$
is completely independent of $\slope$ and the source is perfectly
aligned with the lens ($\srS=0$).

\subsection{Degeneracy between lens profile slope and $\tdist$}
\label{sec:SPL:slopeDegen}

The time delay between images B and A for circularly symmetric surface
mass density follows from equation (\ref{eq:dt}),
\be
\label{eq:dtCircSym}
\dtBA = \frac{\tdist}{c} \left[  \frac{(\imB-\srS)^2}{2}-\psi(\imModB)
  -\frac{(\imA-\srS)^2}{2} + \psi(\imModA) \right ]. 
\ee
For the power-law profile, the following property holds
\be
\label{eq:alpha_psi_reln}
\imMod \alpha(\imMod) = (3-\slope) \psi(\imMod).
\ee
Using equations (\ref{eq:sys1sr}) and (\ref{eq:alpha_psi_reln}) in
equation (\ref{eq:dtCircSym}), we obtain the following relation
between the time-delay distance and model parameters
\be
\label{eq:DdtReln}
\frac{\tdist}{c\,\dtBA} = \frac{2(3-\slope)}{\slope-1}
\left[\imA^2-\imB^2+ \frac{2(2-\slope)}{\slope-1}\srS (\imModA +
  \imModB)  \right]^{-1}.
\ee

\begin{figure}
\centering
\includegraphics[width=0.47\textwidth, clip]{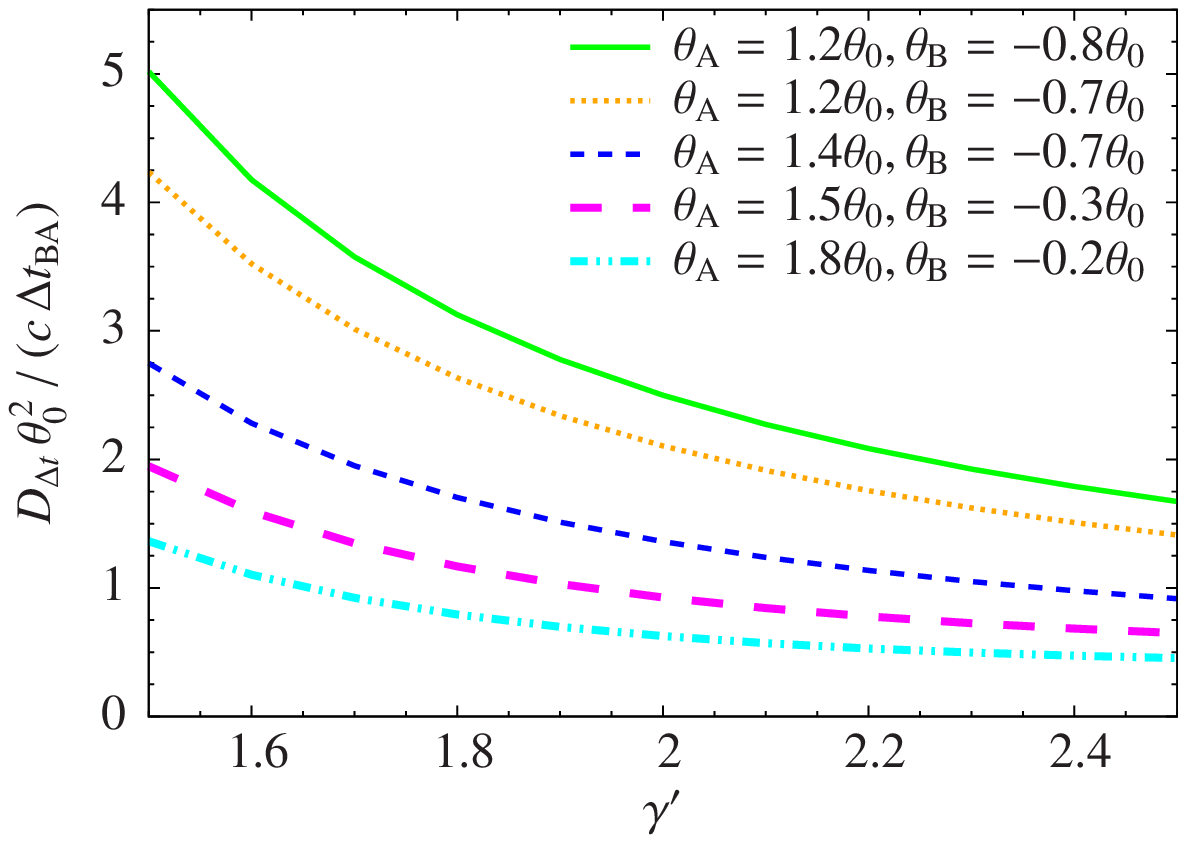}
\includegraphics[width=0.47\textwidth, clip]{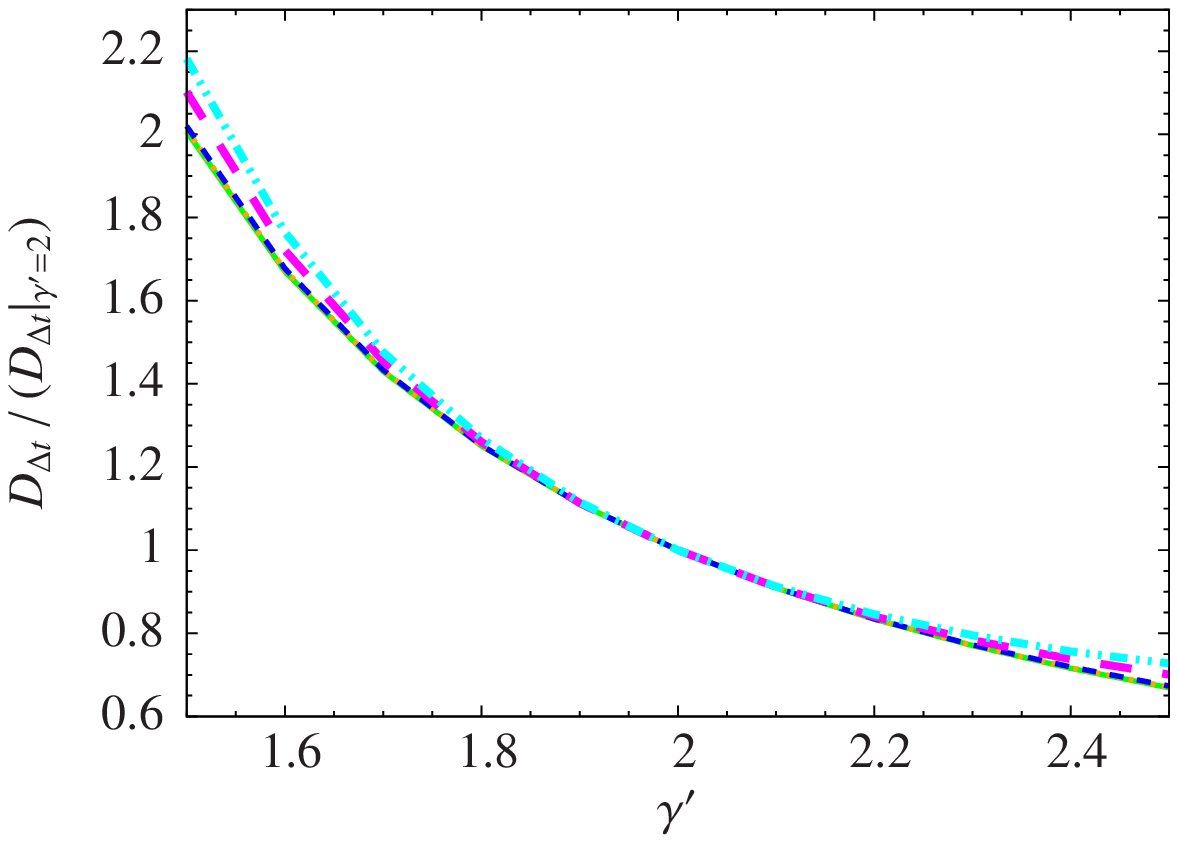}
\caption{\label{fig:Ddt} The time-delay distance as a function of the
  lens profile slope for systems with different image configurations.
  Top panel is 
  the scaled time-delay distance, and the bottom panel is the
  time-delay distance relative to the isothermal ($\slope=2$) case.
  The relative time-delay distance is very similar for all image
  configurations.  An uncertainty in the slope of $0.15$, which is
  roughly the 1-$\sigma$ scatter in profile slopes of SLACS lenses,
  translates 
  to an uncertainty of $\sim$$15\%$ on $\tdist$, hindering
  cosmological studies.}
\end{figure}

This is consistent with equation (22) of \citet{Wucknitz02}.  We
see in equation (\ref{eq:DdtReln}) that $\tdist$ does not scale only
as $\frac{3-\slope}{\slope-1}$ for the spherical power-law model
due to the dependence of the quantities in the square brackets on
$\slope$ (both indirectly via $\srS$ and directly).  In Figure
\ref{fig:Ddt}, we show the time-delay distance scaled by
$\imN^2/c\,\dtBA$ as a function of $\slope$ for Systems II to VI.
System I is not shown since the source is lensed into a ring in this
case so that the time delay between A and B is zero which provides no
constraint on the $\tdist$.  In the bottom panel, we show the
time-delay distance scaled relative to the isothermal case ($\slope=2$).  Lens
systems with different image configurations lead to very similar
relative time-delay distance.  Furthermore, the relative $\tdist$
depends sensitively on $\slope$.  Studies of the SLACS lens galaxies
find a mean slope of $\slope\sim2.07$ with an intrinsic 1$\sigma$ scatter of
$\sim$$0.15$ \citep{AugerEtal10, BarnabeEtal11}.  A change of $\sim$$0.15$
in $\slope$ corresponds to a rescaling of $\tdist$ by $\sim$$15\%$.
Therefore, the isothermal assumption that was frequently imposed in
previous studies of two-image lenses would easily lead to a biased $\tdist$
determination at the $10\%-20\%$ level.  For precision cosmology,
one must therefore measure accurately the slope of the lens mass
profile.  As seen earlier, systems with single-component sources cannot
be used to constrain the slope via image positions alone (equations
(\ref{eq:sys1sr}) are underdetermined).  In the next section, we
consider sources with two components.

\subsection{Lens systems with two-component sources }
\label{sec:SPL:2sr}

In this section, we explore the constraints on the lens profile slope
in systems where the source has two compact components.  We label the image
positions as $\imAi$ and $\imBi$ for the source component at position $\srSi$,
and as $\imAii$ and $\imBii$ for the source component at position $\srSii$.
Note that the two source components need not be collinear (in
projection) with the lens centre.  The four image positions lead to
four constraint equations 
\be
\label{eq:sys2sr}
\left\{ \begin{array}{l}
\srSi = \imAi - \thEin \frac{\imAi}{\imModAi} \left( \frac{\thEin}{\imModAi}\right)^{\slope-2} \\
\srSi = \imBi - \thEin \frac{\imBi}{\imModBi} \left( \frac{\thEin}{\imModBi}\right)^{\slope-2} \\
\srSii = \imAii - \thEin \frac{\imAii}{\imModAii} \left( \frac{\thEin}{\imModAii}\right)^{\slope-2} \\
\srSii = \imBii - \thEin \frac{\imBii}{\imModBii} \left( \frac{\thEin}{\imModBii}\right)^{\slope-2} \\
\end{array} \right.
\ee
With four equations and four unknowns ($\srSi$, $\srSii$, $\thEin$ and
$\slope$), the value of $\slope$ can in principle be solved in the
above system of equations (except for the special case where the two
source components are located equidistant from the lens centre, so
that the first two equations and the last two equations in
(\ref{eq:sys2sr}) are equivalent up to a sign change).  

We now explore the precision in which the
image positions need to be measured in order to determine $\slope$
of the lens to a few percent precision for cosmography.  For
illustrative purposes, we focus on lens system II in Table \ref{tab:toySys}
and consider a range of possible image positions for the second
component in the source that we assume to lie in the same direction
from the lens centre as the first source component.  In particular, we
consider a range of values for $\imAii-\imAi$ spanning from $-0.4\imN$ to
$0.2\imN$.

\begin{figure}
\centering
\includegraphics[width=0.47\textwidth, clip]{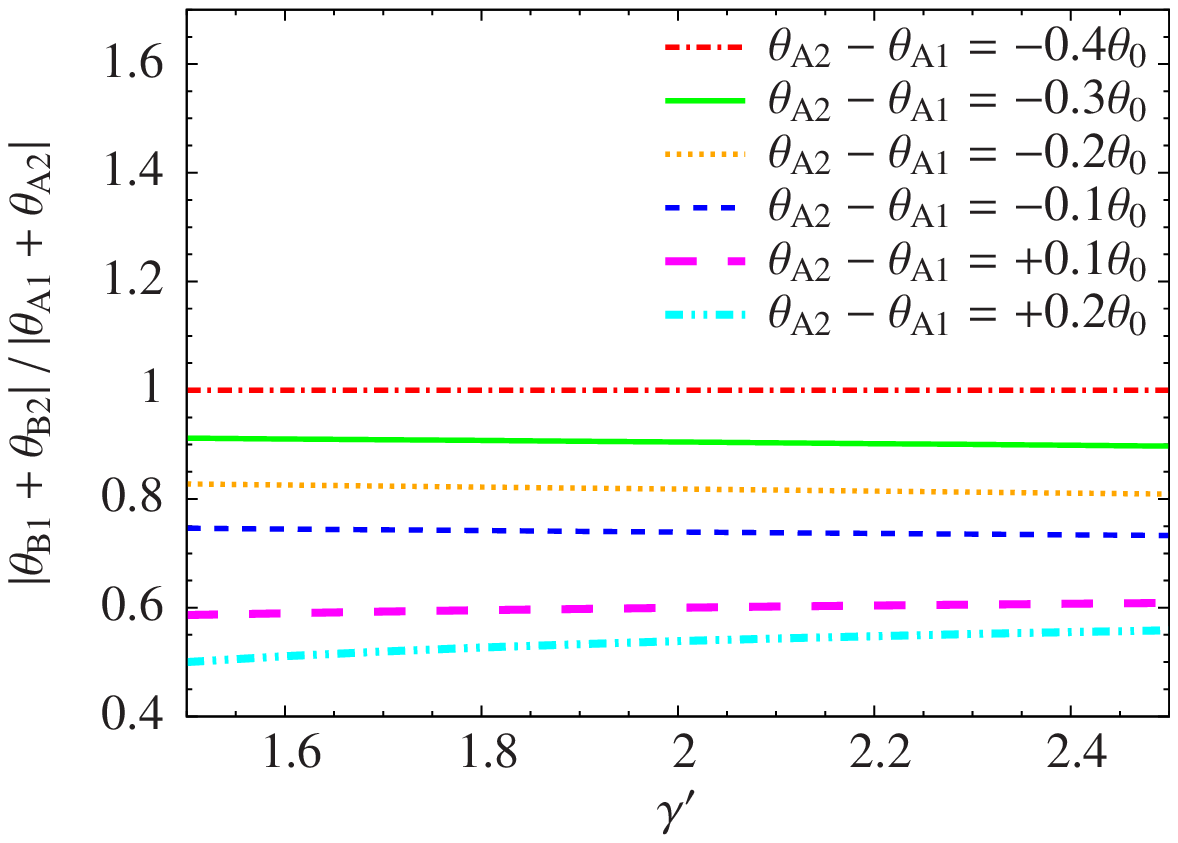}
\includegraphics[width=0.47\textwidth, clip]{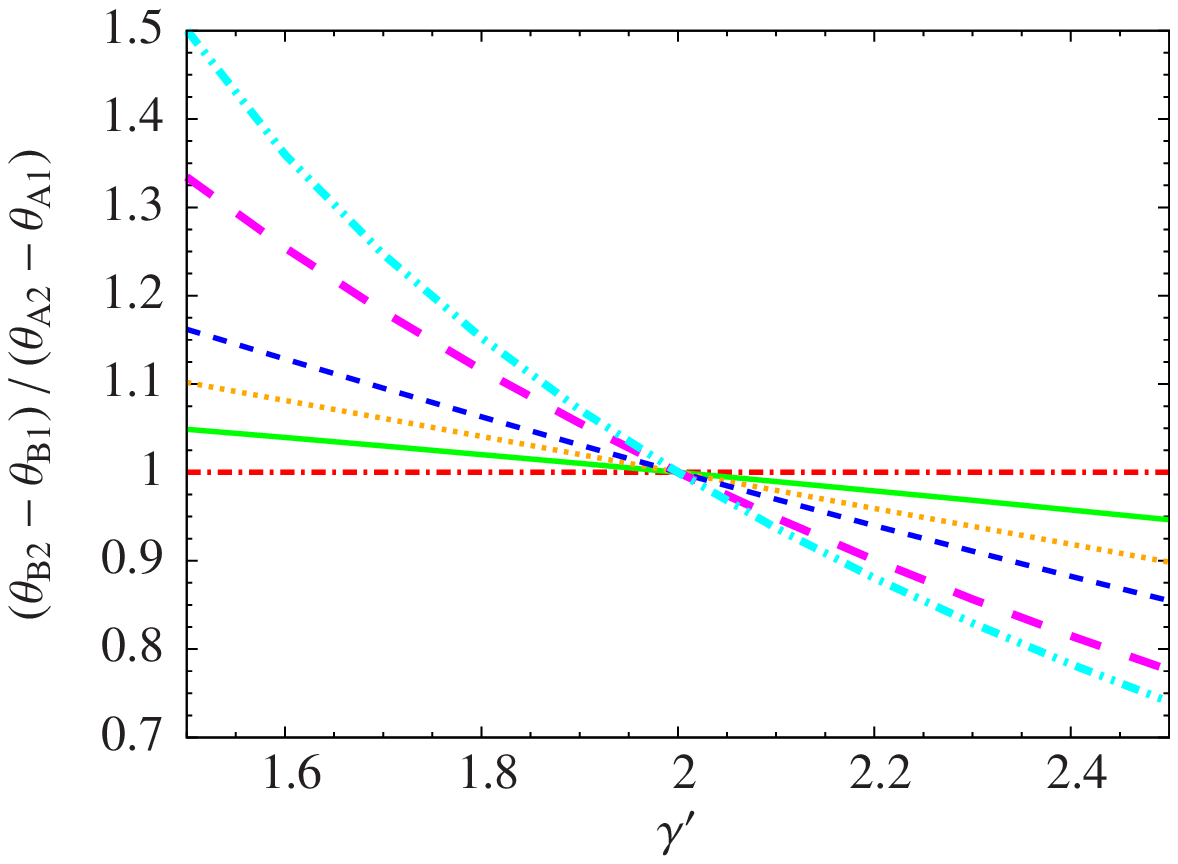}
\caption{\label{fig:B2B1A2A1} Configurations of images with
  2-component sources.  Top panel: asymmetry in the images, as defined
  by the ratio of the average image position of the two components.
  The ratio is 1 for perfect symmetry and is smaller for more asymmetric
  systems.  Bottom panel: ratio of the separations between the images
  of the two components.  For systems that are not perfectly symmetric,
  the curves are strictly monotonic, indicating that the slope can be
  determined in principle by measuring the image separations of the
  components.}
\end{figure}

The top panel in Figure \ref{fig:B2B1A2A1} quantifies the asymmetry in the image
configurations relative to the lens galaxy for each of the
$\imAii-\imAi$ values.  The closer the ratio of the
average image position (for the two-component images) is to 1, the
more symmetric is the image configuration.  For the case where
$\imAii-\imAi=-0.4\imN$, there is perfect symmetry with $\imAi =
-\imBii$ and $\imAii= -\imBi$ so that
$\frac{|\imBi+\imBii|}{|\imAi+\imAii|}=1$ for all $\slope$.  In the
bottom panel of Figure \ref{fig:B2B1A2A1}, we plot the relative image
separation between the two image components as a function of $\slope$.
Apart from the case with $\imAii-\imAi=-0.4\imN$ (red dot-dashed
lines), the derivative of the curves with respect to $\slope$ is
negative.  Therefore, by measuring 
the relative separation between the images of different source
components, one can measure $\slope$ of the lens galaxy.  The
system with $\imAii-\imAi=-0.4\imN$ provides no information on 
$\slope$ because in this perfectly symmetric image configuration, the
second source component is on the opposite side and equidistant from
the lens centre as the first source component, yielding effectively
only a single component source in terms of constraints
(which is insufficient for determining $\slope$, as shown in Section
\ref{sec:SPL:1sr}).

\begin{figure}
\centering
\includegraphics[width=0.47\textwidth, clip]{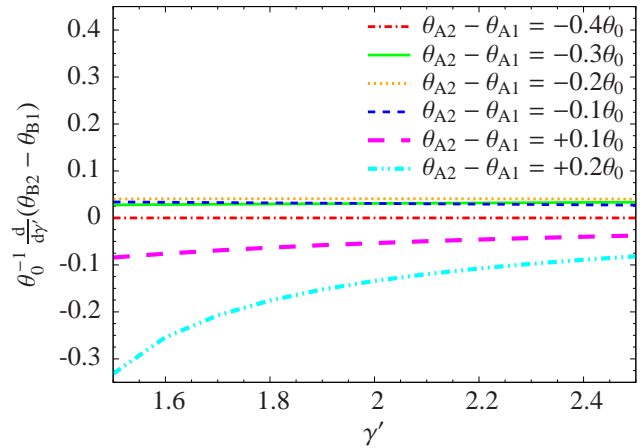}
\caption{\label{fig:dB21dgam} The derivative of the image separation
  between the two components ($\imBii-\imBi$) with respect to the
  slope $\slope$.  Highly asymmetric systems have larger magnitudes
  for the derivatives.  To measure $\slope$ to within $\sim$0.03 for
  precision cosmology, mas precision measurements in the relative
  image separation are required.}
\end{figure}

In Figure \ref{fig:dB21dgam}, we show the derivative of $\imBii-\imBi$
with respect to $\slope$.  For typical systems with
$\frac{|\imBi+\imBii|}{|\imAi+\imAii|}\gtrsim0.7$ that are not
perfectly symmetric, the derivative is $\sim$$0.04\imN$.  Systems that are
highly asymmetric ($\frac{|\imBi+\imBii|}{|\imAi+\imAii|}\lesssim0.7$)
can have larger magnitudes of $\sim$$0.1\imN$ for the derivatives.  In
order to measure $\Delta \slope$ to 
within 0.03 (which translates to $\sim$$3\%$ in $\tdist$), one would
need to measure $\imBii-\imBi$ with accuracies better than
$\sim$$0.1\imN\cdot0.03 = 0.003\imN$.  For typical galaxy-scale lenses with
$\imN\sim1''$, this requires $\lesssim$$3\,$milliarcsecond (mas)
precision measurements on 
the separation between image components.  Therefore, even though
lenses with two-component sources can in principle be used to
constrain the slope of the lens profile, in practise it would be
difficult to measure the image separation between the components with
mas precision to constrain $\tdist$ with a few percent precision.
Image positions of AGNs can be measured with mas precisions using
radio telescopes, but usually the second source components (if any)
are spatially extended radio jets whose positions are typically
measured with precisions of several mas at best.  There are,
  however, radio jets with compact knots that provide mas
  astrometries \citep[e.g.,][]{PatnaikEtal95}.

We have so far considered a power-law profile to describe the lens
mass distribution.  However, mass structures along the line of sight
from us to the source typically induces shear on the system, which is
characterised by a strength and an angle.  Determining the
external shear strength and angle in addition to the two power-law
parameters ($\thEin$ and $\slope$) and the two source positions
($\beta_{\rm s1}$ and $\beta_{\rm s2}$) is not possible since the
images from the two-component sources provide only four constraints.
Nonetheless, for typical two-image systems where the images lie on
opposite sides with a small angular offset (i.e., no longer collinear
with the lens galaxy) due to the presence of a general quadrupole
(including external shear), the time delay depends weakly on the
structure of the quadrupole \citep{Kochanek02}.  Therefore, the
dependence of $\tdist$ on $\slope$ remains roughly the same in the
presence of shear, but $\slope$ becomes much more difficult to
determine. 

Systems with even more source components would provide additional
constraints on the lens mass distribution (e.g., $\slope$
and the external shear), but most systems do not
have multiple compact source components with mas astrometries.
Nonetheless, source galaxies (i.e., the hosts of the AGNs) are 
typically spatially extended, which can be thought of as many point
sources with different intensities.  The lensed images of the
spatially extended sources form arcs, and the relative thickness of
the arcs at various locations helps to constrain $\slope$, just
like the relative separation between $\imBii-\imBi$ and
$\imAii-\imAi$, if measured accurately, constrains $\slope$.  While
the arc thickness cannot be measured with mas precisions at a
particular location even on current high-resolution imagings,
the arc thickness can be measured at many angular positions.  In the
next sections we explore two-image lens systems with extended sources
for cosmological studies.


\section{Simulations}
\label{sec:sim}

We simulate deep and high-resolution imaging of two-image lens systems
that reveals the lensed arcs of the extended source galaxy.  In
particular, we simulate data that mimic the system \helens\
\citep{WisotzkiEtal93} with a lens redshift of 0.729
\citep{LidmanEtal00} and a source redshift of 2.319
\citep{WisotzkiEtal93, SmetteEtal95}.  The corresponding time-delay
distance for the lens is $\tdist=4829\,{\rm Mpc}$ assuming a flat
$\Lambda$CDM universe with $H_0=70\,{\rm km\,s^{-1}\,Mpc^{-1}}$ and
$\Omega_{\Lambda}=0.72$.  The time delay between the images is
$162\pm6$ days \citep{MorganEtal08}.

\subsection{Input lens mass profile and source light profile}
\label{sec:sim:lenssr}

To create simulated images and time delays for the mock systems, we
use an elliptical power-law profile for the lens mass distribution
with a constant external shear.  The form of the elliptical power-law
surface mass density that we employ is 
\be
\label{eq:PLkappa}
\kappa_{\rm epl}(\im_1,\im_2) = \frac{3-\slope}{1+q_{\rm
    d}}\left(\frac{\thEin}{\sqrt{\im_1^2+\im_2^2/q_{\rm d}^2}} \right)^{\slope-1},
\ee
where $q_{\rm d}$ is the axis ratio of the elliptical isodensity contours, and
$\thEin$ is the Einstein radius for the spherical-equivalent case (in
the limit where $q_{\rm d}=1$, the above $\kappa_{\rm epl}$ distribution reduces to
equation (\ref{eq:kappa:thE})).  The deflection angle and lens potential
can be computed following \citet{Barkana98}.  The distribution is
suitably translated to the position of the lens galaxy
($\boldsymbol{\theta_{\rm c}}$) and rotated by
the position angle ($\phi_{\rm d}$) of the lens galaxy
(where $\phi_{\rm d}$ is measured counterclockwise from $\im_1$). 

We use the following form for the lens potential of the constant
external shear in polar coordinates $\imMod$ and $\varphi$:
\be
\label{eq:extsh}
\psi_{\rm ext}(\imMod,\varphi) = \frac{1}{2}\gamma_{\rm ext} \imMod^2
\cos2(\varphi - \phi_{\rm ext}),
\ee
where $\gamma_{\rm ext}$ is the shear strength and $\phi_{\rm ext}$ is
the shear angle.  The shear centre is arbitrary since it corresponds
to an unobservable constant shift in the source plane. Note that
$\kappa_{\rm ext}=\frac{1}{2}\nabla^2\psi_{\rm ext}$ is zero. The
shear position angle of $\phi_{\rm ext}=0\degr$ corresponds to a
shearing along the $\im_1$-direction whereas $\phi_{\rm ext}=90\degr$
corresponds to a shearing in the $\im_2$ direction.

For the surface brightness distribution of the AGN host galaxy
in the source plane, we use S{\'e}rsic profiles with S{\'e}rsic index of 1
(corresponding to exponential profiles).  Furthermore, we add a point
source at the centre of the S{\'e}rsic profile to simulate the AGN.

\subsection{Simulated WFC3 observations and Time Delays}%
\label{sec:sim:wfc3}

We simulate \textit{Hubble Space Telescope} (\HST) imaging using the
Wide Field Camera 3 (WFC3) in the infrared (IR) channel since the
source galaxy is typically brighter in the infrared, providing better
contrast with the AGN.  The simulated image pixel size is $0.09''$,
which can be dithered from images with the native pixel size of $0.13''$.

The steps for creating the simulated image are (1) generate an
extended source intensity distribution with a central point source,
(2) lens the source through the power-law and external shear profiles
(described in Section \ref{sec:sim:lenssr}) with parameters tuned to
produce a high-resolution lensed image mimicking \helens, (3)
convolve the lensed image with a subsampled point spread function
(PSF) that is generated 
using the TinyTim software \citep{KristEtal11}, (4) bin the convolved
image to obtain an image pixel size of $0.09''$, and (5) add uniform
Gaussian noise for the background (with $\sigma_{\rm bkgd}^2=1500$
counts, comparable to the level from a few orbits of $\HST$
observations) and Poisson noise for the source.

We consider three input values for the slope: $\slope=1.8$,
$\slope=1.9$ and $\slope=2.2$, and label them as Simulation
\#1, \#2 and \#3, respectively. 
For each input value, the other lens parameters and the point source
position (of the AGN) are adjusted to create systems with the
astrometry and time delay of \helens.  For Simulations \#1 and
  \#2, we adopt a time-delay distance of $4829\, {\rm Mpc}$
  (corresponding to the fiducial $\Lambda$CDM model) and can simulate
  time delays that are close to the observed delay in \helens.  On the
  other hand, simulating a similar time delay with a much steeper slope of
  $\slope = 2.2$ requires a lower $\tdist$ (as illustrated in Figure
  \ref{fig:Ddt}); consequently, we adopt $\tdist=3263\, {\rm Mpc}$ for
  Simulation \#3.  In addition, Simulation \#3 with a steeper mass
  profile has a lower lensing magnification in comparison to the other two
  simulations.  Therefore, the intrinsic brightness and the size of
  the extended source in Simulation \#3 are higher than those in the other two
  simulations so that the lensed images from the three
  simulations are similar in terms
  of arc thickness and signal-to-noise ratio.  The position angle of
  the source is arbitrarily chosen to be either $0\degr$ or
  $90\degr$; based on Section \ref{sec:SPL:2sr}, we suspect this
  quantity to be of little importance provided that the source is
  of sufficient spatial extent (for measuring the relative thickness
  of the lensing arcs). 
Table \ref{tab:simInput}
summarises the crucial parameters and simulation outputs.  

\begin{figure}
\centering
\includegraphics[width=0.5\textwidth, clip]{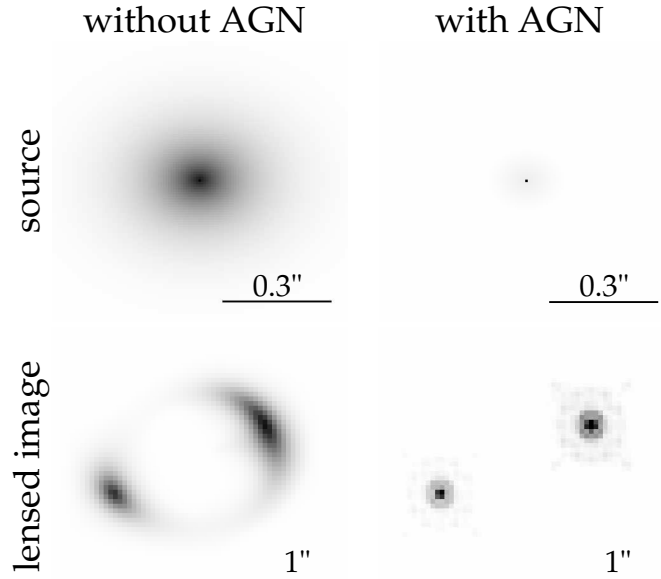}
\caption{\label{fig:simImSr} Simulated \HST\ WFC3 image of Simulation
  \#1.  The left panels show the source galaxy (top) and the lensed
  image without noise (bottom) in the absence of the AGN in the
  source/host galaxy.  The lensed image clearly exhibit the spatially
  extended arcs of the lensed AGN host galaxy.  The right panels show
  the source and the lensed image in the presence of the
  AGN, which is typically much brighter than its host galaxy.  We
  model the simulated WFC3 image with the AGN (bottom-right panel) in
  Section \ref{sec:Mod}.}
\end{figure}

For each simulation of the slope, we also consider three different noise
realisations.  Specifically, we use different random number
  seed to generate the uniform Gaussian noise for the background and
  Poisson noise for the source.
In Figure \ref{fig:simImSr}, we show the source and
the WFC3 image of the first realisation of Simulation \#1 in the right panels (top and bottom,
respectively).  The AGN is typically much brighter than the
source/host galaxy, so only the AGN is conspicuous in these panels.
We show in the left panels the source galaxy and the corresponding
lensed image without the AGN to display the underlying extended arc
features of the lensed AGN host galaxy (purely for illustration
purposes without noise added).  We model the simulated WFC3 image
(bottom-right panel) in Section \ref{sec:Mod}. 
For the lensed AGN, we adopt a typical uncertainty of 4\,mas for the
image positions and use $162\pm6$\,days for the time delay $\Delta
t_{\rm AB}$.

\begin{table*}
\caption{Simulation of Systems Resembling \helens}
\label{tab:simInput}
\begin{center}
\begin{tabular}[b]{lccc}
\hline
Parameter & Simulation \#1 & Simulation \#2 & Simulation \#3\\
\hline
$\imVec_{\rm c}$ (arcsec) & $(-0.955,-0.495)$  & $(-0.960,-0.497)$ & $(-0.955,-0.494)$\\
$\thEin$ (arcsec)          & $1.406$            & $1.333$  & $1.462$\\
$q_{\rm d}$               & $0.825$            & $0.805$ & $0.794$ \\
$\phi_{\rm d}$ ($\degr$) & 160 & 25 & 123 \\
$\slope$  & 1.8  & 1.9 & 2.2\\
\hline
$M_{\rm AGN}$  & 19.5 & 19.5 & 19.5 \\
$M_{\rm host}$ & 23 & 23 & 22.5\\
$r_{\rm eff}$ (arcsec)  & 0.2 & 0.2 & 0.4 \\
$q_{\rm s}$    & 0.8 & 0.8 & 0.8 \\
$\phi_{\rm s}$ ($\degr$) & 0 & 90 & 90 \\
\hline
$\imVec_{\rm A}$ (arcsec)  & $\phantom{-}(0.000,-0.010)$  &
$\phantom{-}(0.000,-0.010)$ & $\phantom{-}(0.000,-0.010)$ \\
$\imVec_{\rm B}$ (arcsec)  & $(-2.910,-1.330)$ & $(-2.910,-1.330)$ &
$(-2.910,-1.330)$ \\
$\tdist$ (Mpc)            & 4829              & 4829  &  3263\\
$\Delta t_{\rm AB}$ (days) & $164$             & 165   &  166 \\
\hline
\end{tabular}
\end{center}
\begin{flushleft}
Notes.  Input Lens and source parameters, and the simulated AGN
positions and time-delays.  The first five rows are the centroid
($\imVec_{\rm c}$), strength ($\thEin$), axis ratio ($q_{\rm d}$),
position angle ($\phi_{\rm d}$) and
radial slope ($\slope$) of the power-law mass distribution for
the lens.  The next five rows are the 
magnitude of the AGN ($M_{\rm AGN}$), the magnitude of the source host 
($M_{\rm host}$), the effective radius ($r_{\rm eff}$), axis ratio
($q_{\rm s}$) and position angle ($\phi_{\rm s}$) of the S{\'e}rsic host.
The next four rows are the two lensed 
image positions of the AGN ($\imVec_{\rm A}$ and $\imVec_{\rm B}$),
the time-delay distance ($\tdist$), and the time delay between the
images ($\Delta t_{\rm AB}$).
\end{flushleft}
\end{table*}

\section{Breaking the $\tdist$-slope degeneracy}
\label{sec:Mod}

In this section, we model the simulated images from the previous
section with the aim to recover the time-delay distance for
cosmography. 

\subsection{Lens and source model}
\label{sec:Mod:lenssr}
To predict the image surface brightness and the time delays, we need to
simultaneously model the lens mass distribution and the 
source surface brightness distribution.  
For the lens mass distribution, we use the same power law and external
shear profiles as in equations (\ref{eq:PLkappa}) and
(\ref{eq:extsh}).  For notational simplicity, we collectively denote
as $\parsVec$ the 6 power-law parameters
  ($\boldsymbol{\theta}_{\rm c}$, $\theta_{\rm E}$, $q_{\rm d}$,
  $\phi_{\rm d}$, 
  $\slope$) and the 2 external shear parameters ($\gamma_{\rm ext}$,
  $\phi_{\rm ext}$). 
For the source surface brightness, we model the AGN light separately
from the host to accommodate the large difference in size and
brightness scales.  We choose to model the lensed AGN as
individual points on the image plane instead of a single point on the
source plane.  In the latter case, one can in principle solve for the
predicted image positions and the image fluxes of a point source on
the source plane for given values of the lens mass parameters (via
equation (\ref{eq:lensEq}) and the magnification dictated by the lens
mass model).  However, the observed image positions and especially the
image fluxes of the point source could deviate from the macro (smooth
power-law) model predictions due to substructure, microlensing, time
delay and dust extinction; these effects could not be easily captured
by a model that has the AGN as a point in the source plane.  Since AGN
image fluxes are typically anomalous \citep[e.g.,][]{DalalKochanek02,
  KochanekDalal04}, we model the fluxes of the lensed AGN images
independently, but require that the image positions of the AGN are
consistent with the macro model up to perturbations caused by, for
example, substructures in the lens mass distribution
\citep[e.g.,][]{ChenEtal07}. With the lensed AGN treated as
individual points (before telescope blurring) and a model for the PSF, we
have three parameters to describe each AGN image: position in $\im_1$
and $\im_2$ and an amplitude.  We collectively denote the parameters
for AGN light as $\agnVec$.  
For the spatially extended host of the
AGN, we model its surface brightness on a grid of pixels
\citep{SuyuEtal06}.

We 
express the predicted image of the lensed source as a vector of pixel
intensities, 
\be
\label{eq:dataP}
\dataVec^{\rm P} = \Bmat\Lmat(\parsVec) \esrVec + \sum_{i=1}^{N_{\rm AGN}}
\eagnVec_i(\agnVec),
\ee
where $\Bmat$ is the blurring operator to account for the PSF,
$\Lmat(\parsVec)$ is the lensing operator that maps source intensity
to the image plane, $\esrVec$ is the vector of source pixel
intensities \citep[see, e.g.,][for details]{SuyuEtal06,SuyuEtal09}, 
$N_{\rm AGN}$ is the number of AGN images, and $\eagnVec_i(\agnVec)$
is the vector of image pixel intensities for image $i$ of the AGN.
This description assumes that the PSF is known a prior and is
fixed.  This is true in practise for \HST\ images which have stable
PSFs that can be modelled with sufficient accuracy based on
instrumental setups and field stars.  

To construct the likelihood function for the lens model using the
pixel intensities, we also require an estimate of the intensity
uncertainty at each pixel.  Both the background (including sky and
read noise) and the astrophysical source contribute to the noise in the
intensity pixels.  We have therefore two terms to describe the
variance of the intensity at pixel $i$,
\be
\label{eq:noiseboost}
\sigma_{{\rm pix},i}^2 = \sigma_{\rm bkgd}^2 + f d_i ,
\ee
where $\sigma_{\rm bkgd}$ is the background uncertainty, $f$ is a
scaling factor, and $d_i$ is the image intensity in counts.  In the
modelling, we adopt $\sigma_{\rm bkgd}^2=1500$\, counts (input to the
simulation) which can in practise be measured from a blank region in
the image without astrophysical sources.  The second term in equation
(\ref{eq:noiseboost}), $f d_i$, corresponds to a scaled version of
Poisson noise (with $f=1$ as the usual Poisson noise).  The value of
$f$ is chosen so that the reduced $\chi^2$ is $\sim$1 for the lensed
image reconstruction \citep[see, e.g.,][for details on the computation
of the reduced $\chi^2$]{SuyuEtal06}.  For a model where the PSF and
the mass distribution are known perfectly, the usual Poisson noise
with $f=1$ typically leads to a reduced $\chi^2\sim1$.  In practise,
models are often simplified versions of reality so there could be
residual features in the image reconstruction with a corresponding
reduced $\chi^2$ that is $\gtrsim1$.  These residuals are frequently
most prominent at locations where the intensities are high, such as at
the positions of the AGN images.  In this case, the model would try to
reduce the high residual at a few localised locations (i.e., the AGN
positions) instead of fitting to the overall structure of the data
(i.e., the lensing arcs), which could lead to biased estimates of
model parameters that are designed to characterise the global
features.  Increasing $f$ has the effect of downweighting these
localised pixels with high $d_i$ and allowing the model to fit to the
overall structure of the data.  Furthermore, the increased $f$ value
with an associated reduced $\chi^2\sim1$ avoids underestimating the
parameter uncertainty in simple models that are meant to characterise
the large-scale features and not necessarily the small-scale features
in the data \citep[e.g.,][]{BrewerEtal12}. 

In sum, the parameters for our model are $\parsVec$, $\esrVec$ and
$\agnVec$, and for simplicity, we fix the PSF to the input TinyTim
PSF.  Given a set of values for $\parsVec$ and $\agnVec$, the
determination of the source surface brightness of the AGN host is a
linear inversion \citep[e.g.,][]{WarrenDye03, SuyuEtal06,
  VegettiKoopmans09}.  Therefore, 
$\esrVec$ are known as linear parameters, and $\parsVec$ and $\agnVec$
are the nonlinear parameters.  

\subsection{Parameter Sampling and Priors}
\label{sec:Mod:samp}

We model the simulated images and time delay with {\sc Glee}\footnote{a
lens modelling software developed by S.~H.~Suyu and A.~Halkola
\citep{SuyuHalkola10,SuyuEtal11}} that has been enhanced to
incorporate AGN modelling.  Since the AGN dominates the flux in the
images, we first optimised for its position and amplitude while fixing
the host flux to be zero.  We then sample the posterior probability
distribution function (PDF) of nonlinear parameters
$\parsVec$ and $\agnVec$ with Markov chain Monte Carlo (MCMC) methods.
We follow \citet{DunkleyEtal05} for efficient MCMC sampling and for
assessing chain convergence.

Bayes' Theorem states that the posterior PDF is
\be
P(\parsVec,\agnVec | \dataVec, \dt) \propto P(\dataVec,\dt
| \parsVec,\agnVec) P(\parsVec,\agnVec).
\ee
Since the imaging and time-delay data sets are independent, the
likelihood separates:
\be
P(\dataVec,\dt | \parsVec,\agnVec) = P(\dataVec | \parsVec,\agnVec)
P(\dt | \parsVec,\agnVec).
\ee
The likelihood of the simulated WFC3 data, $P(\dataVec
| \parsVec,\agnVec)$, is obtained by reconstructing the AGN host surface
brightness distribution given values of $\parsVec$ and $\agnVec$, and
marginalising over the source parameters $\esrVec$ \citep[see,
e.g.,][for details]{SuyuHalkola10}.  Since the AGN images are modelled
as independent points based on their surface brightness, we include an
additional term to the likelihood 
to ensure that the lens model can reproduce the image positions.  In
particular, we multiply the original likelihood $P(\dataVec 
| \parsVec,\agnVec)$ by 
\be
\prod_{i}^{N_{\rm AGN}} \frac{1}{\sqrt{2\pi}\sigma_i}
\exp\left[-\frac{|\imVec_i-\imVec^{\rm P}_i(\parsVec)|^2}{2\sigma_i^2}
\right],
\ee
where $\imVec_i$ is the measured image position, $\sigma_i$ is the
positional uncertainty that we adopt as 4\,mas, and $\imVec^{\rm P}_i$ is the
predicted image position given 
the lens parameters.  The likelihood for the time delay is given by 
\be
P(\dt | \parsVec,\agnVec) = \frac{1}{\sqrt{2\pi}\sigma_{\dt}} \exp
\left[ {-\frac{(\dt-\dt^{\rm P}(\parsVec,\agnVec))^2}{2\sigma_{\dt}^2}} \right],
\ee
where $\dt$ is the measured time delay with uncertainty $\sigma_{\dt}$
and $\dt^{\rm P}$ is the predicted time delay computed via equation
(\ref{eq:dt}).  

For the prior PDF $P(\parsVec,\agnVec)$, we assume uniform priors for
$\parsVec$ and $\agnVec$ within physical ranges (e.g., axis ratio is
uniform between 0 and 1) except for the centroid of the lens where we
assume a Gaussian prior with width of $0.01''$ centred on the input
values (which in practise would be obtained from the lens light
distribution).  For the pixelated source surface brightness
distribution of the host, we consider both the curvature and
  gradient forms of 
regularisation/prior \citep[see, e.g., Appendix A of ][]{SuyuEtal06}.

\subsection{Recovery of slope and $\tdist$}
\label{sec:Mod:result}
\begin{figure}
\centering
\includegraphics[width=0.5\textwidth, clip]{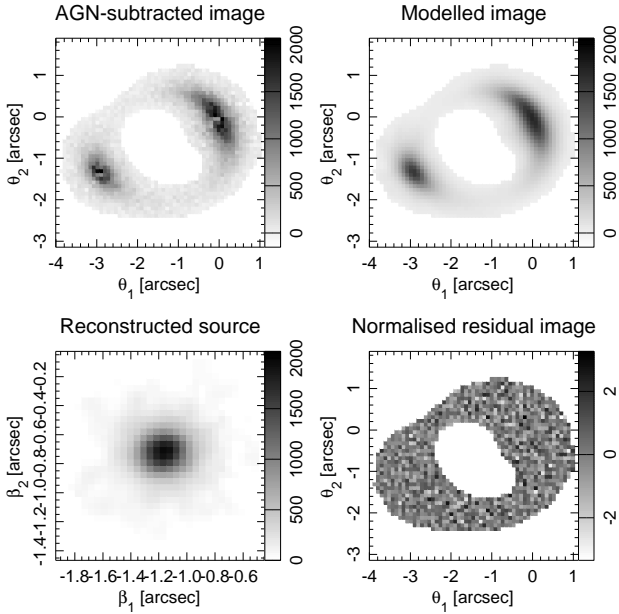}
\caption{\label{fig:reconImSr} Image and source reconstruction of
  the first realisation in Simulation \#1.  Panels from top left in clockwise
  direction: image with AGN point images subtracted, modelled image,
  normalised image residual (in units of the uncertainty for each
  pixel), and the reconstructed AGN host galaxy.}
\end{figure}

We model the time delay and the simulated WFC3 image for each of the
three realisations of the three simulations.  To explore the effects of
the different forms of regularisations for the source surface
brightness, we adopt the curvature form for Simulations \#1 and
\#2, and the gradient form for Simulation \#3.
In Figure \ref{fig:reconImSr}, we show the source and image
reconstruction of the AGN host surface brightness for the most
probable lens and AGN parameters in the first realisation of
Simulation \#1.  The noise level near the core of the lensed images is high
due to the Poisson noise from the AGN.  The modelled source resembles
the input source in Figure \ref{fig:simImSr} and reproduces the
corresponding simulated lensed image.

\begin{figure*}
\centering
\includegraphics[width=0.95\textwidth, clip]{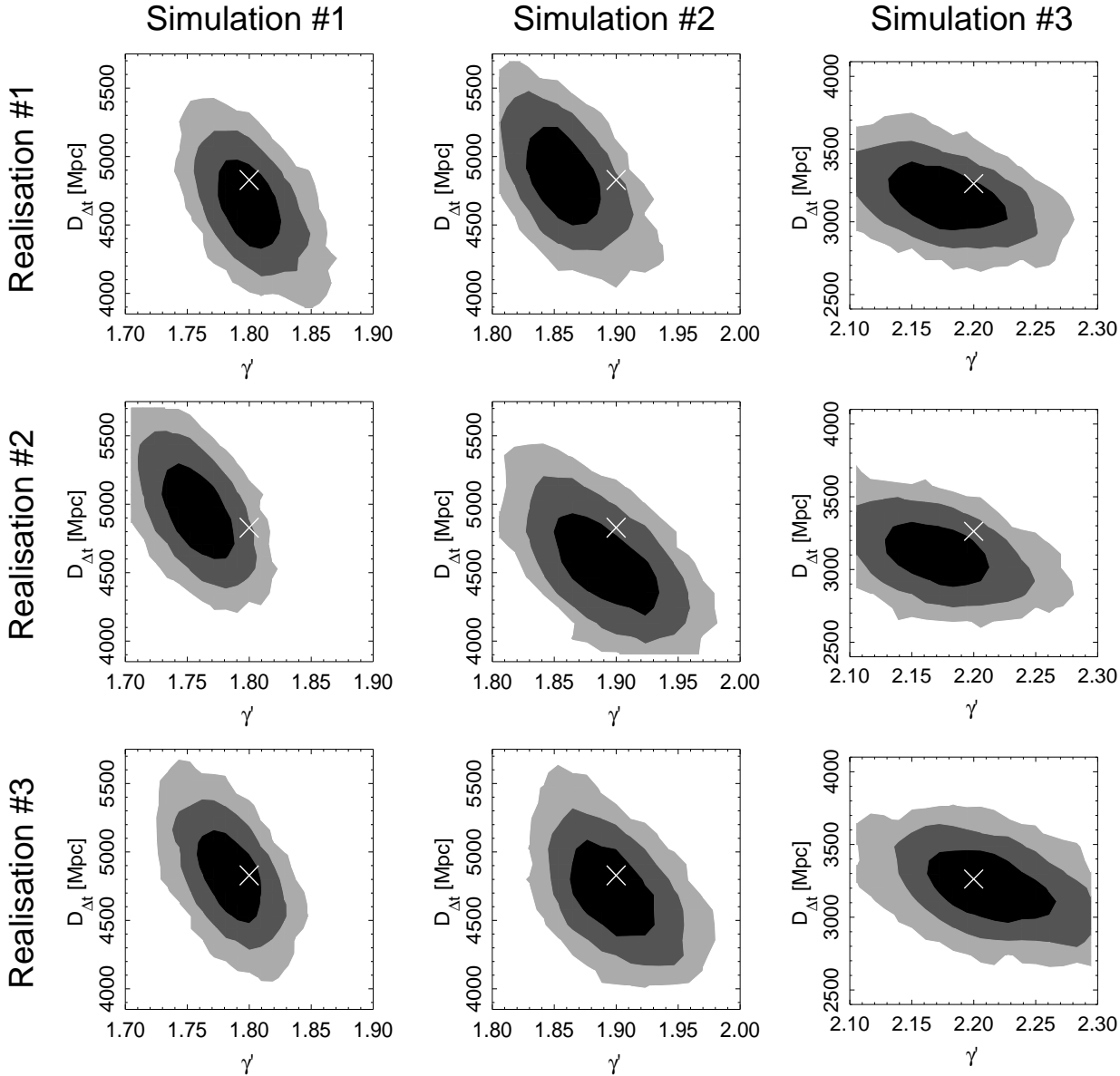}
\caption{\label{fig:PDF_Dtgam} Marginalised PDF of $\slope$ and
  $\tdist$ for the three simulations (columns) with three
    realisations (rows) each.  
  The three shaded areas show the 68.3\%, 95.4\% and 99.7\% credible
  regions.  The crosses mark the input values.
  By using the extended surface
  brightness of the AGN host, we break the degeneracy between 
 $\tdist$ and $\slope$ and recover the input $\tdist$ value within the
uncertainties.  Two-image lens systems with imaging that reveals
the AGN host provide a robust cosmological distance probe.}
\end{figure*}

In Figure \ref{fig:PDF_Dtgam}, we show the joint PDF for $\slope$ and
$\tdist$ after marginalising over all other parameters (source pixel
intensities, AGN positions and amplitudes, power-law mass parameters and
external shear) in each of the simulations/realisations.  The cross
indicates the input parameter value. 
The shapes and sizes of the credible regions are similar for the different
realisations in each simulation.  The orientations of the credible
regions follow the degeneracy curves in Figure \ref{fig:Ddt}, and
become more horizontal for simulations with higher $\slope$.  
We recover the true $\slope$ and $\tdist$ values in Table
  \ref{tab:simInput} within the 99.7\% (3$\sigma$) credible region for
all cases, and within the 68.3\% (1$\sigma$) credible region for
approximately 2/3 of the cases as expected.  By using the extended surface
brightness of the host, we break the $\tdist$-$\slope$ degeneracy and
recover $\tdist$ to within $5\%$.  Therefore, even though the width of
the lensing arc cannot be measured to mas accuracy at any particular
location given the image pixel size of $0.09''$, the thousands of image
pixels collectively determine the relative thickness of the lensing
arcs to sufficient accuracy for cosmography.

\subsection{Discussions}
\label{sec:Mod:disc}

Our simulations demonstrate that two-image systems with detectable
extended images of the AGN host provide accurate constraints on the
time-delay distance, overcoming the radial profile slope degeneracy
that has undermined this cosmological probe in the past.  Only
four-image lens systems or two-image lens systems with multi-component
source structures (which have more observational constraints) have so far
been shown to yield cosmological measurements that are not obviously
dominated by systematic effects \citep[e.g.,][]{CourbinEtal11, SuyuEtal10,
  FadelyEtal10, WucknitzEtal04}.  Nonetheless, the majority of
currently known time-delay lenses are two-image systems without
multi-component sources \citep[e.g.,][]{Oguri07, ParaficzHjorth10}.
Furthermore, future telescopes will find $\sim$$6$ times more
two-image time-delay systems than four-image systems, and the Large
Synoptic Survey Telescope (LSST) will discover thousands of two-image
time-delay lenses \citep{OguriMarshall10}. Therefore, effectively
tapping into the abundant reservoir of two-image lenses will provide
significant advances to time-delay cosmography.

The profile slope of interest for cosmography is actually the slope in
the annulus between the images since the time delay primarily depends
on the average surface mass density between the images
\citep{Kochanek02}.  Therefore, even if the true lens mass
distribution is not a global power law, it is well approximated by a
local power law at the positions of the images.  Other methods have
probed the mass distribution in regions outside of the image annulus;
for example, stellar kinematics of lens galaxies further constrain the
mass distribution inside the effective radius of the lens galaxy
\citep[e.g.,][]{KoopmansTreu03, TreuKoopmans04, KoopmansEtal09,
  BarnabeEtal09, AugerEtal10, BarnabeEtal11, SonnenfeldEtal11}.  In
fact, stellar kinematics also help break the so-called ``mass-sheet
degeneracy'' \citep{FalcoEtal85} in lensing
\citep[e.g.,][]{GroginNarayan96a, KoopmansEtal03, SuyuEtal10}.  Mass
structures along the line of sight between the observer and the source
(such as individual galaxies or groups/clusters of galaxies)
contribute an external convergence, $\kext$, to the lens mass
distribution, and this external convergence is degenerate with
$\tdist$.  Specifically, there is a mathematical transformation to the
lens mass distribution, $\kappa\rightarrow (1-\kext)\kappa+\kext$,
which leaves the lensing observables (e.g., image
positions/morphology, relative image fluxes and time delays) invariant
but rescales $\tdist$.  A model that does not account for an existing
$\kext$ would underpredict or overpredict the value of $\tdist$ for an
overdense or underdense line of sight, respectively (by a factor of
($1-\kext$)).  Qualitatively, having a nonzero external convergence is
analogous to adding an extra lens to the system which changes the
focal length of the system, and hence the distance
measurement.  Both the radial profile slope degeneracy and the
mass-sheet degeneracy are known to be the dominant sources of
systematic uncertainties in measuring $\tdist$.  We have tackled and
eliminated the radial profile slope degeneracy in this paper.  Lens
environment studies \citep[e.g.,][]{KeetonZabludoff04,
  FassnachtEtal06, MomchevaEtal06, FassnachtEtal11} in conjunction
with stellar kinematics are effective in breaking the mass-sheet
degeneracy \citep{SuyuEtal10}.

Our modelling of the lensed images requires a good knowledge of the PSF
that is stable.  Space-based imaging is ideal, but existing
\textit{HST} archival images of two-image lens systems that are not in
cluster environments (which make $\kext$ hard to control) do not have
sufficient signal-to-noise ratio in the extended images of the AGN
host galaxy for accurate modelling.  Furthermore, most lens systems
have been imaged with the Near Infrared Camera and Multi-Object
Spectrometer (NICMOS) that have non-linear count rates; modelling
these images, which have intensities of the extended AGN host galaxy
that are severely 
contaminated by the bright AGNs, is prone to systematic effects.
In contrast, WFC3 is a more sensitive and stable detector with a wider
field of view (hence more field stars for PSF models) that would
currently be the optimal instrument on \textit{HST} to follow up the
two-image lens systems for cosmography.

\begin{figure}
\centering
\includegraphics[width=0.45\textwidth, clip]{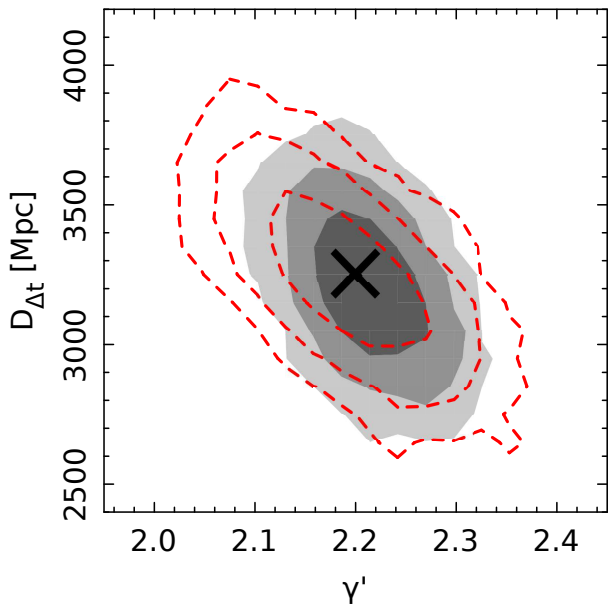}
\caption{\label{fig:PSFmismatch} The effect of the PSF.
    Marginalised PDF of $\slope$ and $\tdist$ for Simulation \#3
    Realisation \#3 modelled using the input TinyTim PSF (shaded) and
    a TinyTim PSF that is offset by $\sim$$45''$ from the input PSF
    (dashed).  The three areas show the 68.3\%, 95.4\% and 99.7\%
    credible regions.  The cross marks the input values.  With the
    imperfect (offset) PSF and the scaled pixel uncertainty in 
    equation (\ref{eq:noiseboost}), the precision of the recovered
    $\tdist$ is slightly degraded due to misfits near the bright AGN
    images.}
\end{figure}

We have assumed in the model that the PSF is known perfectly, which is
not true in practise.  To explore the effect of imperfect PSF
knowledge, we have also generated another TinyTim WFC3 PSF, located at
approximately $45''$ from the original PSF, and used this offset PSF
in the modelling step instead of the original PSF.  This corresponds
to the scenario where a star in the field is used to approximate the
PSF at the location of the lens, which has been shown to work well for
lens systems without bright AGNs in the spatially extended sources
\citep[e.g.,][]{MarshallEtal07, SuyuEtal09}.  In our case where the
source AGN is bright, the PSF mismatch leads to imperfect AGN image
modelling and consequently significant images residuals near the
positions of the bright AGNs.  
As a result, the scaling factor $f$ in equation (\ref{eq:noiseboost})
is $\sim$8 for the offset PSF (whereas $f=1$ for the perfect PSF) to
obtain a reduced image $\chi^2$ of $\sim$1 by downweighting these
bright pixels with residuals that could otherwise cause biases in
parameter estimations.  Figure \ref{fig:PSFmismatch} shows the
resulting constraints on $\tdist$ and $\slope$ (dashed) by using the
offset PSF with the scaled pixel uncertainty in equation
(\ref{eq:noiseboost}).  The input $\tdist$ and $\slope$ are recovered
without significant biases.  
In comparison to the case with a perfect PSF (shaded), the recovery of
$\tdist$ is degraded in precision from $\sim$$5\%$ to $\sim$$6\%$
(1$\sigma$, after marginalising over other parameters) due to the
downweighting of high-residual pixels and the consequent loss of
information in these pixels.

For the simulated WFC3-IR images, we find that subsampling is
necessary to avoid biases in the recovered parameters due to the
large image pixel sizes.  In other words, the predicted lensed image
first need to be created and convolved on a finer resolution, then
binned to the observed image resolution for calculating the
likelihood.  A subsampling factor of $\sim3$ was sufficient to
characterise both the PSF and the source intensity variation in the
simulations.  

In modelling the simulated images, a positional uncertainty of 4\,mas
was adopted for the AGN images and a Gaussian prior with width of
$0.01''$ was imposed on the lens galaxy centroid.  We consider the
impact of relaxing these constraints individually to $0.09''$ (1
pixel) for observations where the positions cannot be easily
measured (e.g., the lens galaxy is faint).  We find that the credible
regions in Figure \ref{fig:PDF_Dtgam} remain nearly the same if either
the positional uncertainty of the AGN images or the prior on the lens
galaxy centroid is relaxed to $0.09''$.  This shows that the spatially
extended arcs are providing most of the constraints on the mass
distribution and the time-delay distance.

We have considered two types of source regularisations for the
simulations (curvature for Simulations \#1 and \#2, and gradient for
Simulation \#3), and showed that the input $\tdist$ and $\gamma'$ are
recovered irrespective of the choice in regularisation (Figure
\ref{fig:PDF_Dtgam}).  For a given simulation, the two forms of
regularisations lead to similar shapes and sizes of credible
regions with slight shifts that are small compared to the size of the
regions.  Therefore, both forms of regularisations are viable options
in modelling the extended arcs for typical AGN host galaxies that have
smooth surface brightness distributions.  

We have kept our simulations simple by excluding the light from the
lens galaxy.  In practise, the observed image would also contain lens
light, which would affect the arc light and would need to be modelled as
well.  One way is to use S{\'e}rsic profiles to describe the lens light,
and add another term to the right-hand side of equation
(\ref{eq:dataP}) for the vector that describe the lens light
intensities.  The parameters for the S{\'e}rsic profiles are also
nonlinear (like $\parsVec$ and $\agnVec$) and need to be sampled as
well.  A self-consistent model would simultaneously determine the lens
light, the lens mass distribution, the AGN contribution, and the AGN
host source surface brightness distribution.  This is a
high-dimensional nonlinear problem that is beyond the scope of this
paper, and will be presented in a future study.


\vspace{-0.2cm}
\section{Conclusions}
\label{sec:conclude}

We have used the spherical power-law model to show that the time-delay
distance depends sensitively on the slope, which cannot be constrained
with a single-component compact source.  A change in the slope of
$\Delta\slope\sim0.15$, which is the typical scatter in lens galaxy
slopes, leads to a $\sim15\%$ change in $\tdist$, undermining the use
of two-image systems for accurate cosmological studies.  Systems with
two-component sources can be used to constrain the slope and derive
useful cosmological constraints, but the relative separation between
the images of the two components need to be measured with mas
precision, which is difficult in practise.  We use simulated \HST\
images to test the usefulness of two-image systems with spatially
extended arcs of the lensed AGN host galaxy.  By simultaneously
modelling the AGN light contribution, the lens mass profile, and the
extended AGN host surface brightness distribution, we find that the
relative thickness of the arcs accurately constrains the lens mass
distribution and results in robust recovery of $\tdist$ to a few
percent.  By establishing that two-image systems are no longer
hindered by the radial profile slope degeneracy, the sample of useful
time-delay lenses is enlarged by a factor of $\sim$$6$ which will
provide substantial advances for cosmological studies.


\vspace{-0.6cm}
\section*{Acknowledgments}

SHS thanks T.~Treu, M.~Auger and O.~Wucknitz for useful discussions
and encouragement.  SHS is grateful to the anonymous referee for the
helpful comments that improved the presentation of the paper.


\vspace{-0.6cm}
\bibliography{ms}
\bibliographystyle{mn2e}


\bsp
\label{lastpage}
\end{document}